\documentclass[vecphys]{svmult}

\usepackage{makeidx}         
\usepackage{graphicx}        
\usepackage{multicol}        
\usepackage[bottom]{footmisc}

\makeindex             

\begin{document}

\title*{Near-field Cosmology with the VLT}
\author{Steffen Mieske\inst{1}\and
Helmut Jerjen\inst{2}}
\institute{ESO Garching, Karl-Schwarzschild-Str. 2, 85748 Garching, Germany
\texttt{smieske@eso.org}
\and Mt\,Stromlo Observatory,\,Australian National University \texttt{jerjen@mso.anu.edu.au}}
%
%
\maketitle With the arrival of wide-field imagers on medium-size
telescopes (e.g.~SkyMapper, Pan-STARRS) and the future co-existence 
of LSST with the E-ELT, it is worthwhile to evaluate the scientific potential 
of a CCD camera with $\approx$ 1 degree FOV for the VLT. Here we 
discuss the role that such an instrument could play in resolving persisting 
fundamental problems in ``near-field cosmology".


\section{Science case}
Dwarf galaxies in the local universe are easily studied survivors from the epoch of galaxy
formation, and thus preferred targets to establish empirical benchmarks for
high redshift cosmological studies, in particular calibrating theories of
galaxy formation and interpreting the observed galaxy luminosity function.  While
$\Lambda$CDM predictions for structure formation on large scales agree
well with the distribution of baryonic matter (galaxies), it is in the
low mass regime where strong discrepancies persist between the 
expected frequency of low-mass dark matter halos and the
number of known dwarf galaxies (e.g. \cite{Moore99},
~\cite{Klypin99}, \cite{Trenth02b}, \cite{Mieske07}). The most prominent place 
affected by this so-called substructure crisis is the Local Group, even
when accounting for the recently discovered ultra-faint dwarf
spheroidals (\cite{Simon07}, \cite{Walsh07}).

A solution of this problem may well be found in fundamental physics such as 
warm or self-interacting dark matter (\cite{Bode01}, \cite{Sperge00}).
Alternatively, there is no shortage of astrophysical mechanisms that
can diminish the accumulation of baryons in low mass dark matter potential wells:
long cooling times for primordial gas in small halos (\cite{Haiman96}), galactic winds driven by supernovae and hot stars (\cite{Dekel86}), or pressure support against collapse of the intergalactic
plasma after reionization (\cite{Thoul96}, \cite{Gnedin00}).

The present picture is further confused because there are a multitude of
reasons why the faint end of the galaxy luminosity function, the optical manifestation
of the dark matter mass function and completely governed by {\it dwarf} galaxies (\cite{Bingge88}), 
could deviate from the simple CDM theory expectation.  Progress on this fundamental issue 
is currently limited by observations, not theory as most recent studies of the galaxy
luminosity function in Virgo (\cite{Philli98}, \cite{Trenth02a}, \cite{Sabati05})
and Fornax (\cite{Kambas00}, \cite{Hilker03}, \cite{Mieske07}) demonstrated that {\it
the ambiguity in attributing membership status to cluster/group galaxy candidates} is the 
prime source of uncertainty on the quest to find the accurate shape, slope, and 
possible turning point of the galaxy luminosity function.


This ambiguity can only be resolved by deriving genuine distances to complete 
populations of dwarf galaxies. 




\section{Galaxy distances from surface brightness fluctuations}
\begin{figure}
\centering
\includegraphics[height=6cm]{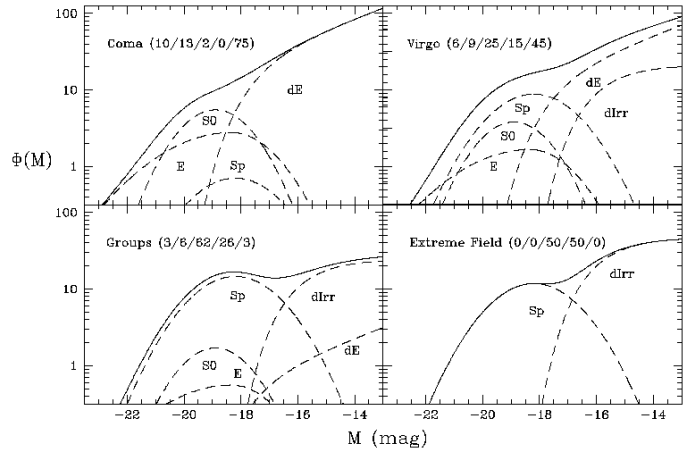}

%
%
\caption{Galaxy luminosity functions for a range of environments, broken down 
into different morphological types (Jerjen 2000). The faint end of the luminosity function 
in dense environments is dominated by early-type dwarf elliptical galaxies.}
\label{fig:1}       
\end{figure}

Especially in dense environments, the faint end slope of the 
galaxy luminosity function is completely determined by the large number of dwarf elliptical galaxies, stellar 
systems primarily composed of old stars and having an almost featureless morphology (Fig.~\ref{fig:1}).
An intriguing possibility to directly derive distances to such
galaxies is the Surface Brightness Fluctuation (SBF) method, whose
theoretical framework was developed by \cite{Tonry88}. The method quantifies the statistical pixel-to-pixel
variation of star counts across a galaxy image with the major
technical advantage of working on unresolved stellar populations.
Since these variations normalised to the underlying mean galaxy light
are inversely proportional to distance (see Fig.2), the SBF amplitude
can be used as a distance indicator, once the dependence of the
amplitude on stellar content (age, metallicity) is corrected for (e.g.
\cite{Tonry01}).  Jerjen and collaborators (\cite{Jerjen98}, \cite{Jerjen00}, \cite{Jerjen01}, \cite{Rekola05} and
Mieske et al.  (\cite{Mieske03b}, \cite{Mieske07}) demonstrated that
the method works well for low surface brightness dEs as faint as
$\mu_{\rm B, eff} =26$\,mag\,arcsec$^{-2}$ and $M_B= -10$\,mag, out to
distances of 20 Mpc using 8m class telescopes.

\begin{figure}
\centering
\includegraphics[height=4cm]{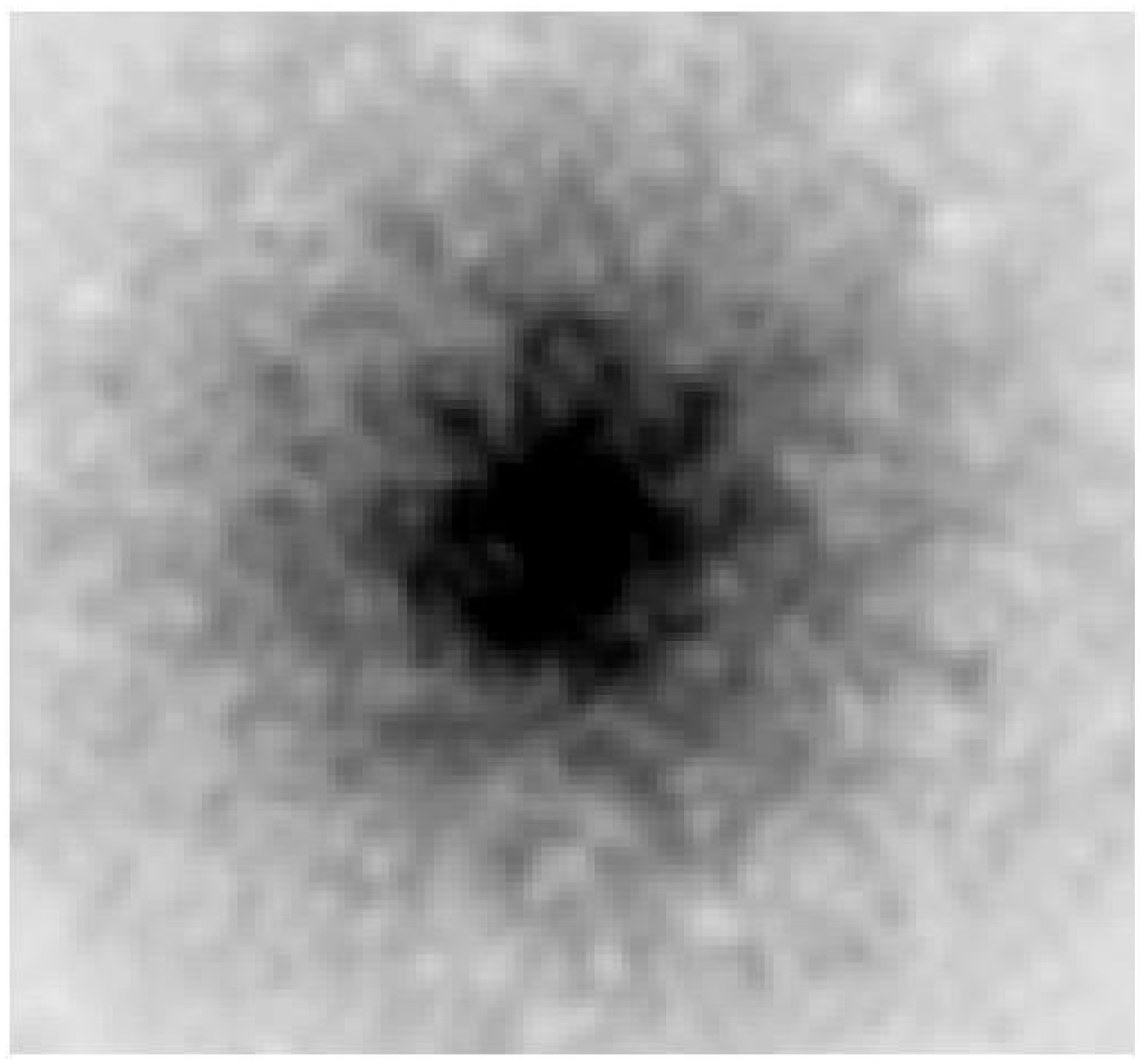}
\includegraphics[height=4cm]{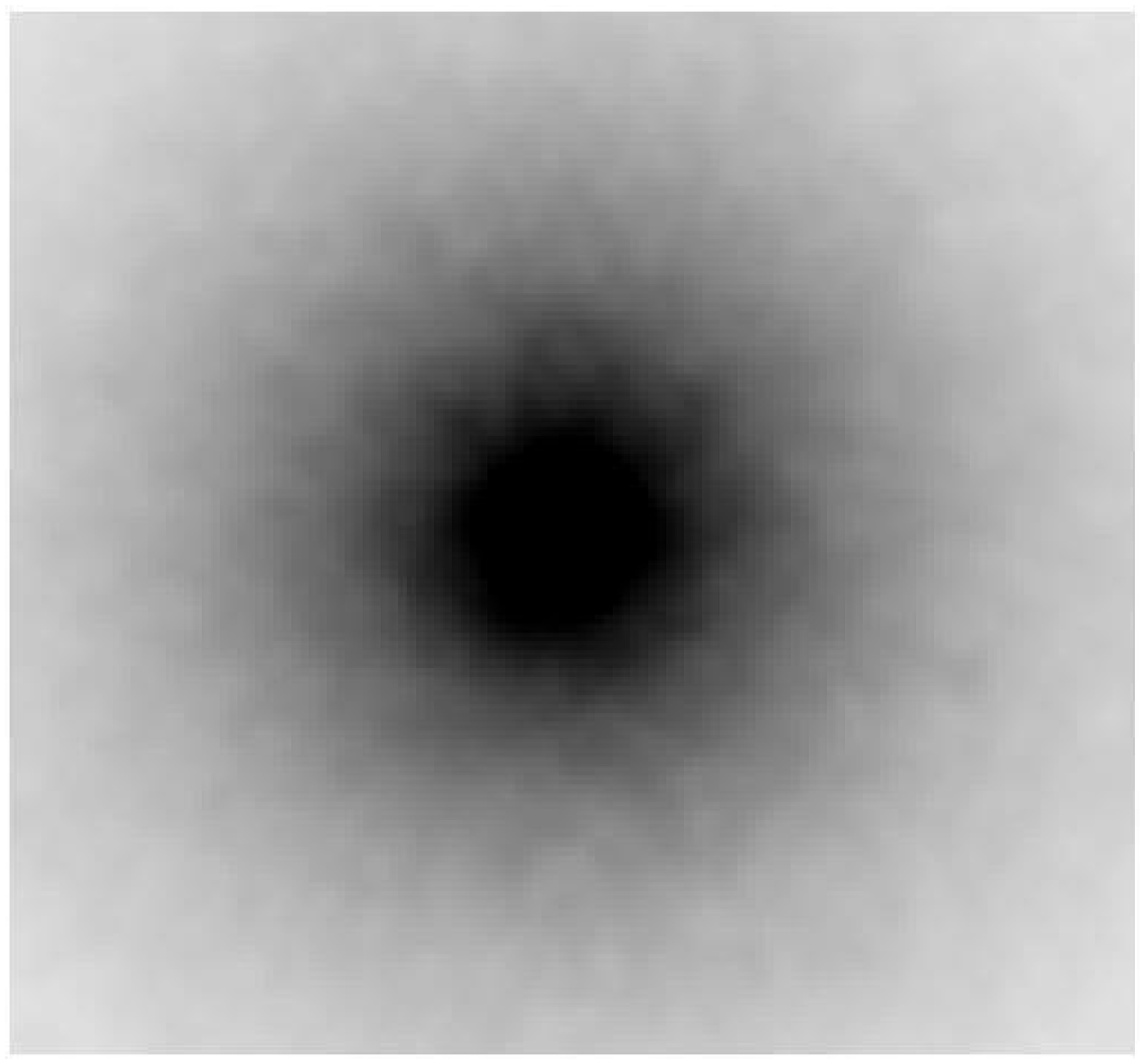}

\caption{The principle of SBF distance estimates. Two simulated images of early-type galaxies with implemented surface brightness fluctuations, having identical angular size but  different distance (\cite{Mieske03a}). 
The galaxy on the right is at a 4 times larger distance than the galaxy on the left giving a smoother apppearance. }
\label{fig:2}       
\end{figure}

\section{Statistical analysis of dwarf galaxy properties}
Beyond 30-40\,Mpc, the SBF method is not efficient anymore for 
measuring distances to faint dwarf galaxies (\cite{Mieske03a}).  
However, near-field cosmology clearly should include the study
of low-mass substructures over a range of environments of different
scale length: inner-cluster distribution, cluster-to-cluster
variations, distribution bias with respect to large-scale filaments
and voids. To achieve a proper sampling of these different levels of
structure scales, one must extend dwarf galaxy studies out to z=0.05
to 0.1 (see for example \cite{Rosenb04}), to cover typical
filament length scales of a few hundred Mpc. This corresponds to an
area on the sky of a few thousand square degrees.

\begin{figure}
  \centering
\includegraphics[height=4cm]{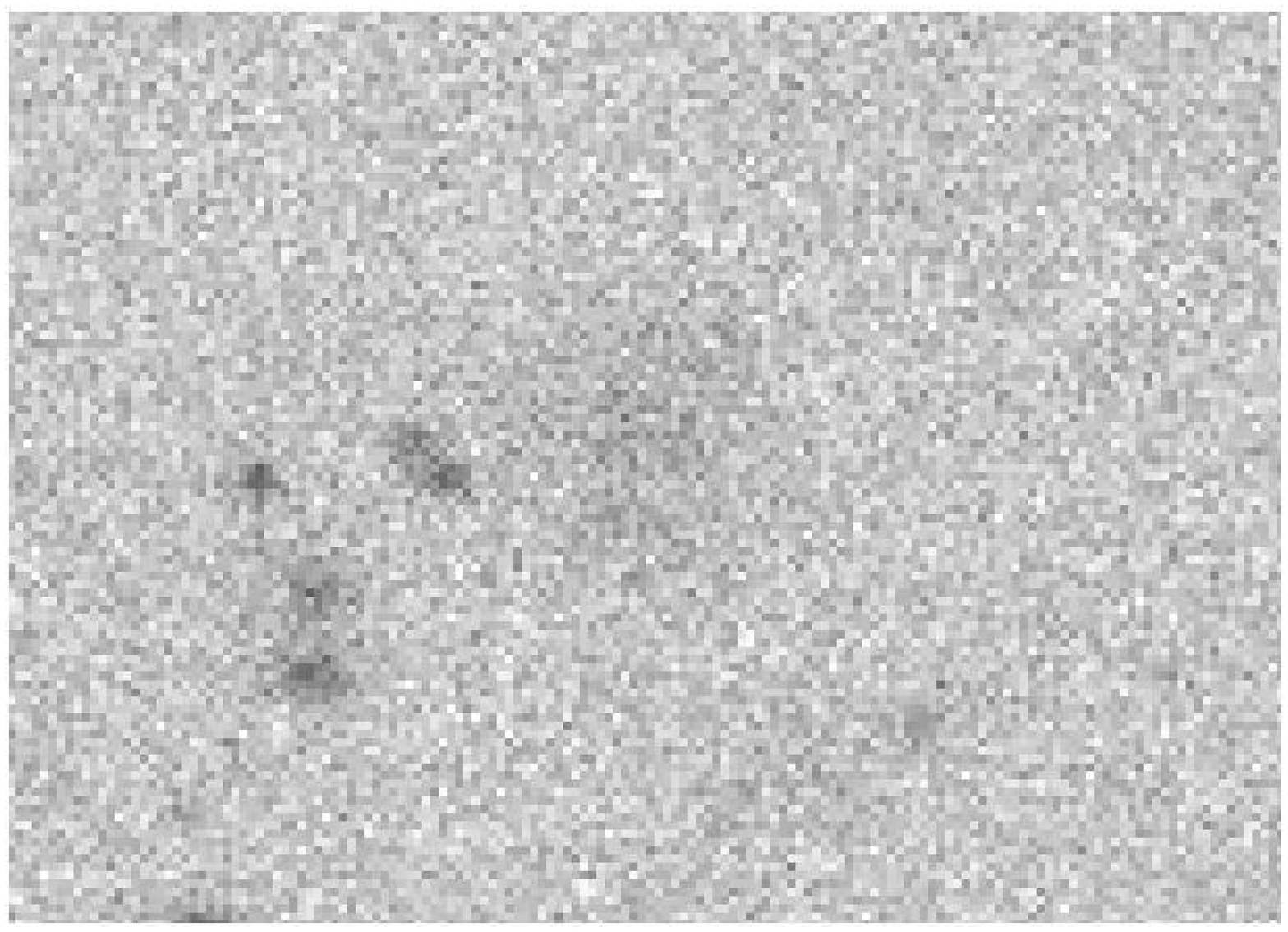}
\includegraphics[height=4cm]{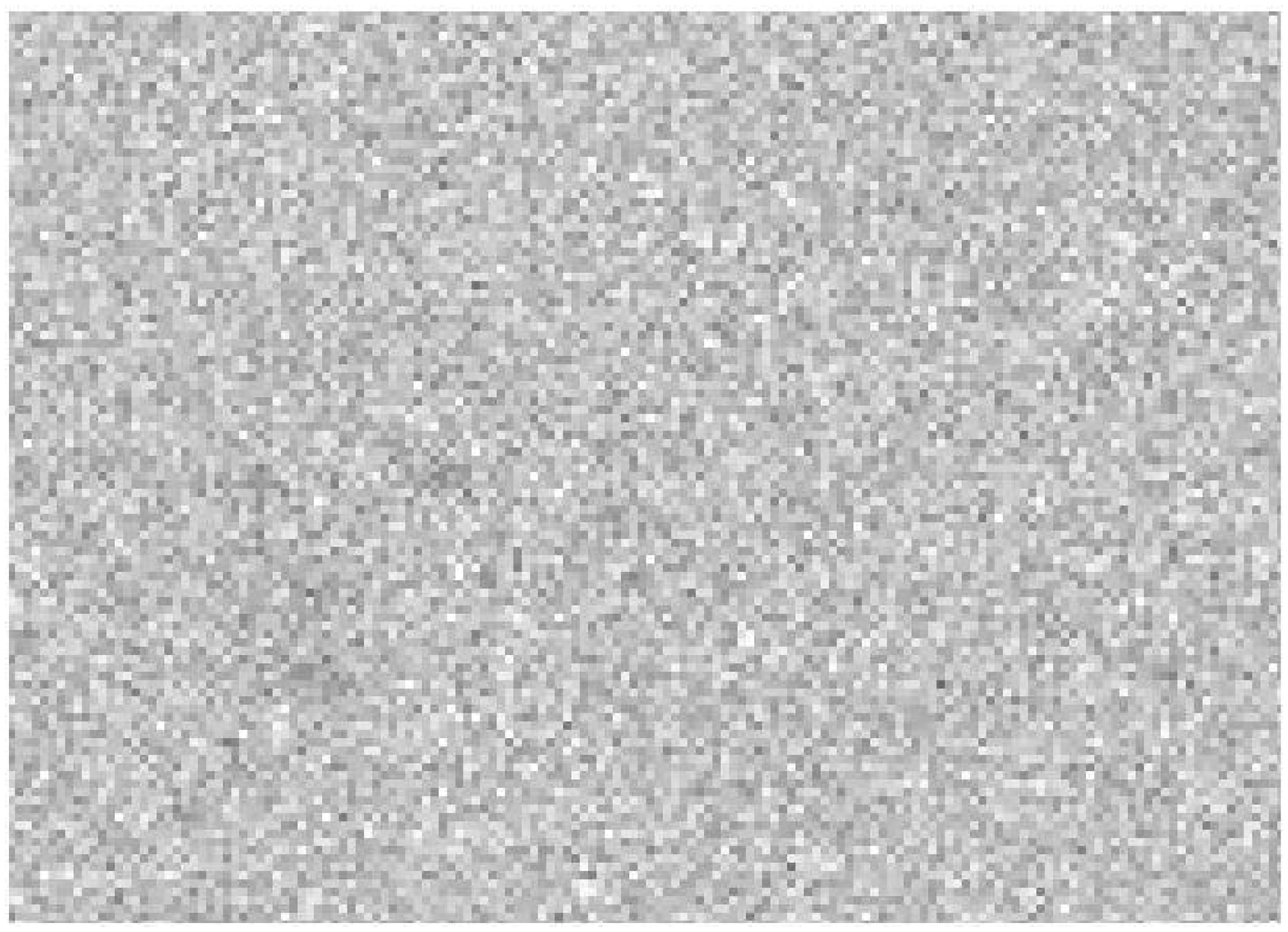}
\caption{Here we compare the image obtained from a 200s exposure
in V of a dwarf galaxy with $M_V=-11$ mag at a distance of 20 Mpc. 
The left image corresponds to the VLT mirror size (8.4m), the right image 
to the VST mirror size (2.6m). The galaxy is only detectable on the VLT image.}
\label{fig:3}       
\end{figure}

From Fig.~\ref{fig:3} it is clear that even with relatively short
exposures of a few minutes on an 8m class telescopes, low surface
brightness galaxies with $M_V \sim -11$ mag can be detected.
Considering the decrease of angular size with distance, more realistic
detection and classification limits for dwarf galaxies are $M_V \sim
-11$ out to 50 Mpc, and $M_V \sim -13$ out to 200 Mpc (z=0.05). Going
back to Fig.~\ref{fig:1}, these magnitude limits are well in the regime where
the faint end slope $\alpha$ dominates the shape of the galaxy
luminosity function.

Of course, detecting low-surface brightness dwarf galaxy candidates
alone is insufficient to derive their absolute luminosity and constructing 
a fiducial galaxy luminosity function. Therefore, a deeper SBF survey 
is required covering a few control regions like dense clusters and
loose groups to calibrate secondary distance modulus estimators, such 
as angular size vs. central surface brightness, and
colour (e.g. \cite{Mieske07}).

\section{Scientific aims and survey setup}

In summary, we advocate the combination of a smaller scale but deep
imaging survey with a shallow imaging survey over a much larger area,
in order to study the properties of low-mass galaxies in the nearby
universe as a function of environment.
The scientific topics that can be addressed by conducting such a near-field 
cosmology survey are the following:

\begin{itemize}
  
\item Contrast the galaxy luminosity function with the expected
  $\Lambda$CDM mass spectrum.  At which luminosity/mass do baryons
  decouple from dark matter? How does this depend on environment?
  This will give crucial input for studies on dark matter phase space
  properties, reionization, feedback, photoionization.

\item What are the clustering properties, spatial and angular
  correlation function of low mass galaxies compared to $\Lambda$CDM
  predictions. What is the origin of satellite galaxies?

\item Constrain the anisotropy of
  dark matter distribution on scales $\le$100 Mpc (Great Attractor.
\item Morphological segregation/transformation, harrassment, as a function
of environment.
\item Synergy with other scientific areas include the study of
  globular cluster systems, intra-cluster light, far-field cosmology
  surveys: ISW, cluster counting, weak lensing, BAO.

\end{itemize}

While the shallow, dE identification survey could in principle also be done based on data
from future surveys with 4m class telescopes, the deep distance survey requires 
the light collecting area of an 8m class telescope. This in order to directly derive 
distances to faint dwarf galaxies with the SBF method.  For the deep survey we 
estimate an area of about 500 square degrees would be sufficient to fully cover the 
most prominent nearby galaxy clusters (d$\le$50 Mpc) and a substantial portion of the low
density field environment. It would require about one hour of total integration
time per pointing shared between V and I (or B and R) band exposures,
in order to derive reliable SBF distances for d$\le$50 Mpc and $M_V \simeq -14$ mag (\cite{Mieske03a}).  Most of the time will be used for the
red filter, which is generally better suited for SBF measurement (\cite{Tonry01}, \cite{Jerjen01}, \cite{Mieske03b}, \cite{Mieske05}, \cite{Mieske06}, \cite{Mei05}).  The
colour information is used to correct the SBF amplitude for stellar
population effects. 

With a wide-field imager (1 sq degree FOV) at the VLT, one would require about
60 nights of observing time for the deep survey. Assuming a 5000 sq degree 
coverage for the shallow survey, one would require roughly 100 nights of 
observing time for 200 seconds exposures in two optical filters V and I (or B and R).


\label{sec:1}

%
%



\printindex

\begin{thebibliography}{}

\bibitem{Bingge88}Binggeli, B., Sandage, A., \& Tammann, G.A. 1988, {ARAA} 26, 509 
\bibitem{Bode01}Bode, P., Ostriker, J.P., \& Turok, M. 2001, ApJ, 556, 93
\bibitem{Dekel86}Dekel, A., \& Silk, J. 1986, {ApJ} 303, 39
\bibitem{Gnedin00}Gnedin, N.Y. 2000, {ApJ} 542, 535
\bibitem{Haiman96}Haiman, Z., Thoul, A.A., \& Loeb, A. 1996, {ApJ} 464, 523
\bibitem{Hilker03}Hilker, M., Mieske, S., \& Infante, L. 2003, A\&AL, 397, L9
\bibitem{Jerjen98}{Jerjen, H., Freeman, K.C., Binggeli, B.} 1998, {AJ} 116, 2873 
\bibitem{Jerjen0} Jerjen, H., Encyclopedia of Astronomy and Astrophysics, ed.~Paul Murdin, 
Bristol: Institute of Physics Publishing, 2000
\bibitem{Jerjen00}{Jerjen, H., Freeman, K.C., Binggeli, B.} 2000,  {AJ} 119, 166 
\bibitem{Jerjen01}Jerjen, H. et al. 2001,  {A\&A} 380,90  
\bibitem{Kambas00}Kambas, A., Davies, J. I., Smith, R. M., Bianchi, \& S.,
Haynes, J. A., 2000, AJ 120, 1316
\bibitem{Klypin99}Klypin, A., Kravtsov, A. V., Valenzuela, O., \& Prada, F. 1999, ApJ, 522, 82
\bibitem{Mei05}Mei, S. et al. 2005, ApJ, 625, 121
\bibitem{Mieske03a}Mieske, S., Hilker, M. \& Infante, L. 2003a, A\&A, 403, 43
\bibitem{Mieske03b}Mieske, S. \& Hilker, M. 2003b, {A\&A} 410, 445
\bibitem{Mieske05}Mieske, S., \& Hilker, M. \& Infante, L. 2005, A\&A, 438, 103
\bibitem{Mieske06}Mieske, S., Hilker, M. \& Infante, L. 2006, A\&A, 458, 1013
\bibitem{Mieske07}Mieske, S., Hilker, M., Infante, L., Mendes de Oliveira, C. 2007, A\&A, 463, 503
\bibitem{Moore99}Moore, B., Ghigna, S., Governato, F. et al. 1999,  ApJL, 524, 19
\bibitem{Philli98}Phillipps, S. et al. 1998, ApJ, 493, L59
\bibitem{Rekola05}{Rekola, R., Jerjen, H. \& Flynn, C.} 2005, {A\&A} 437, 823
\bibitem{Rosenb04}Rosenbaum, S.D., \& Bomans, D. 2004, A\&A, 422L, 5
\bibitem{Sabati05}Sabatini, S. et al. 2005, MNRAS, 357, 819
\bibitem{Simon07}Simon, J. D., \& Geha, M. 2007, ApJ, 670, 313 
\bibitem{Sperge00}Spergel, D.N. \& Steinhardt, P.J. 2000, Physical Review Letters 84, 3760
\bibitem{Thoul96}Thoul, A.A. \& Weinberg, D.H. 1996, {ApJ} 465, 608
\bibitem{Tonry88}{Tonry, J.L. \& Schneider, D.P.} 1988, {AJ} 96, 807
\bibitem{Tonry01}{Tonry, J.L. et al.} 2001, {ApJ} 546, 681
\bibitem{Trenth02a}Trentham, N. \& Hodgkins, S. 2002, {MNRAS} 333, 423 
\bibitem{Trenth02b}Trentham, N., \& Tully, R.B., 2002, MNRAS 335, 712
\bibitem{Walsh07} Walsh, S.M., Jerjen, H., Willman, B., 2007, ApJL, 659, 121
\end{thebibliography}
\end{document}